\documentclass[prd,nofootinbib,12pt]{revtex4}

\usepackage{lineno,hyperref}
\usepackage{amsfonts,amssymb,amsmath}
\usepackage{epsfig}
\usepackage{mathrsfs}
\usepackage{amssymb}
\usepackage{graphicx}
\usepackage{amsmath}
\usepackage{graphicx}
\usepackage{amsmath}
\usepackage{xcolor}
\usepackage{color}

\begin{document}

\title{The running curvaton}

\author{Lei-Hua Liu$^{1}$}
\email{liuleihua8899@hotmail.com}
\author{Wu-Long Xu$^{2}$}
\email{wlxu@emails.bjut.edu.cn}

\affiliation{$^1$Department of Physics, College of Physics, Mechanical and Electrical Engineering, Jishou University, Jishou 416000, China}
\affiliation{$^2$Institute of Theoretical Physics, Beijing University of Technology, Beijing 100124, China}

\begin{abstract}
Inspired by \cite{Jiang:2018uce}, we propose a similar curvaton mechanism whose realization occurs in preheating process, in which the effective mass is running (its potential consists of coupling part and exponential part whose contribution is subdominant comparing to the coupling part). The production of curvaton contains the cases of narrow resonance and broad resonances whose criteria comes via the spectral index of curvaton. Since the inflationary potential is chaotic inflation (quadratic potential), it could smoothly transit into the preheating process. Once the entropy perturbation transferred into curvature perturbation, we will use $\delta N$ formalism to investigate its validity. By neglecting the contribution of exponential potential of curvaton, we calculate power spectrum $P_\zeta$ and non linear Non-Gaussian parameter $f_{NL}$. Our calculation analytically shows that these two observables are independent of potential of inflaton. Finally, as the curvaton almost decay (inflaton field vanishes), the exponential potential will be approaching a constant of order of cosmological constant, which may play a role of dark energy.

\end{abstract}

\maketitle


\section{Introduction}

In the mainstream of inflationary theory, the origin of curvature perturbation is generated by the inflationary perturbation seeding the CMB temperature fluctuations. However, this broad kind of theories highly depends on the shape of inflationary potential. In order to relax this strict condition, many alternatives are proposed. Such kind of models called curvaton mechanism are given, in which the curvaton field is an extra scalar field and is subdominant during inflation. After inflation, the density of curvaton will be more and more significant which will generate the curvature perturbation (much larger than the curvature perturbation from inflaton)~\cite{Enqvist:2001zp, Moroi:2001ct, Lyth:2001nq}.

All of the particles of the standard model (SM) come from preheating or reheating period \cite{Traschen:1990sw,Kofman:1994rk,Shtanov:1994ce,Prokopec:1996rr,Greene:1997ge,Kofman:1997yn,Greene:1997fu}. Especially for the narrow resonance, Ref.~\cite{Shtanov:1994ce} gives the general discussion under the framework of the WKB approximation. Since the reheating temperature is not sufficient for the coupling between the inflaton and other particles of SM, the preheating process is mandatary \cite{Allahverdi:2010xz} in some specific inflationary models. Once adding the curvaton decay after inflation, most of them decay perturbatively~\cite{Bartolo:2002vf}. It also has been envisaged that curvaton could non-perturbatively decay \cite{Enqvist:2008be}. Furthermore, the curvaton can be dubbed as a source to generate amount of gravitational waves (GW) during preheating~\cite{Figueroa:2017vfa}. Due to the rich phenomenon of curvaton, it could be embedded into multi-field framework \cite{Liu:2019wrb}.

Curvaton could couple to the Higgs field in which case the mass of the curvaton can vary significantly~\cite{Enqvist:2012tc,Enqvist:2013gwf}. In this paper, we propose the similar idea, which the coupling is between the inflaton and curvaton, to vary the mass of curvaton. Generally, curvaton decay occurs when the curvaton decay rate $\Gamma_\chi$ is comparable to Hubble parameter $H$. Upon relaxing the condition that the condensate of curvaton dominates over its perturbations, it could yield large local non-Gaussianities~\cite{Lyth:2002my}. Current observations~\cite{Ade:2015ava} severely constrain these models however, as (local) non-Gaussianity $f_{NL}$ cannot be too large ($|f_{NL}|< 10$), it rules out curvaton models that produce large non-Gaussianities. However, this local type of Non-Gaussianity can be suppressed by quadratic potential plus quartic potential \cite{Mukaida:2014wma} and also in string axionic potential \cite{Dimopoulos:2011gb,Kawasaki:2012gg}. Once taking the local Non-Gaussianity into account, we will see that it will strongly constrain the decay of convaton, precisely for the so-called fraction of curvaton's energy density among radiation period, the newest investigations can be found in \cite{Sharma:2019qan}.

Curvaton field is generally deemed as an independent and extra scalar field comparing to inflaton. Thereby, curvaton could play various roles in particles, such as axion \cite{Gong:2016yyb}, in order to account for dark matter (DM). Due to the role of axion, curvaton could also produce the axionic-primordial blackhole \cite{Kawasaki:2012wr,Ando:2018nge}, in some sense it explains the DM. Furthermore, primordial blackhole as DM could be generated by curvaton and inflaton mixed model \cite{Chen:2019zza}.

In a traditional scenario of curvaton, the curvature perturbation will be generated after inflation since the curvaton lives longer than inflaton. Meanwhile, the main contribution of curvature perturbation comes from curvaton comparing to inflaton, which is considered as an assumption for curvaton mechanism. In our model, the curvaton field comes via the decay of inflaton, in which it means that there is a direct coupling between the curvaton field and inflaton field. For this scenario, we consider the potential of curvaton contains the coupling part (dominant part) and exponential part (mimicking the dynamical behavior of dark energy). Once the curvaton decays into other particles ($\it i.e.$ Higgs particles, $W^{\pm}$ $\it e.t.c$), the potential of curvaton will be approaching a constant deemed as playing a role of cosmological constant. In some sense, our scenario could account for the origin of dark energy from the perspective of phenomenology. Therefore, we give a fully analysis of curvaton from its generation in an inflationary period to the very late Universe (up to the present dark energy epoch).

This paper is organized as following. In section~\ref{model} we introduce our inflationary model with two scalars,
one being inflaton and the other is curvaton, whose potential (curvaton's potential) contains two part: coupling part (between the curvaton and inflaton), exponential part. In section~\ref{production of curvaton}, the production of curvation comes via the decay of inflaton in terms of parametric resonance preheating. In section \ref{power spectrum}, the detailed calculation of power spectrum and its corresponding local Non-Gaussianity are given by $\delta N$ formalism. In section \ref{dark energy epoch}, we illustrate that the dark energy comes from the exponential potential of curvaton from the perspective of phenomenology. Finally, section \ref{conclusion} gives the conclusion.

We work in natural units in which $c=1=\hbar$, but retain the Newton constant $G$.


\section{The model}
\label{model}

The (pre)reheating process provides an environment for generating the particles, meanwhile, it also produces the entropy perturbation. One essential ingredient of realizing the preheating process is the so-called parametric resonance, requiring that there is a coupling between the inflaton field and other field. Since mechanism of (pre)reheating is rather crude, it remains lots of heuristic places for investigating.

In order to achieve the curvaton mechanism under the framework of preheating, Ref. \cite{Jiang:2018uce} numerically study the curvature perturbation produced by the entropic field, which could be regarded as a particular realization of curvaton mechanism in preheating process. In light of this realization, we could construct a direct curvaton scenario.
Subsequently, in order to account for the origin of dark energy, we assume that the second part of the potential for curvaton is of exponential form. Therefore, the total action can be constructed as following,
\begin{equation}
S=\int d^{4}x\sqrt{-g}\bigg\{\frac{M_{P}^{2}}{2}R-\frac{1}{2}g^{\mu\nu}\nabla_{\text{\ensuremath{\mu}}}\phi\nabla_{\nu}\phi-\frac{1}{2}g^{\mu\nu}\nabla_{\text{\ensuremath{\mu}}}\chi\nabla_{\nu}\chi
-V(\phi)-\frac{g_0}{M_{P}^{2}}\chi^{2}V(\phi)-\lambda_{0}\exp[-\lambda_{1}\frac{\chi}{M_{P}}]\bigg\}
\label{total action}
\end{equation}
where $\chi$ and $\phi$ denote curvaton and inflaton, respectively, $R$ presents the Ricci scalar and $g$ is the determinant of $g_{\mu\nu}$. $g_0$, $\lambda_0$ and $\lambda_1$ are the dimensionless parameters determined by observations in Lagrangian.

In order to get more understanding of this scenario, we elaborate action (\ref{total action}) a bit more. In some sense, the curvaton could produce via inflaton decay \cite{Byrnes:2016xlk} albeit the branch of decay cannot be large. In what following, we will show how the curvaton produce via the parametric resonance in preheating process.

\section{Preheating of production of curvaton}
\label{production of curvaton}
In order to produce the curvaton field, it will be generated by the parametric resonance. Generally, the production of curvaton contains two parts: one is the background considered as a classical field, the other is quantum fluctuations of curvaton. For simplicity, we will concentrate the main contribution of background field of curvaton assuming that only depends on time.

\begin{figure}[h!]
 \centering
  \includegraphics[height=8.8cm, width=8.52cm]{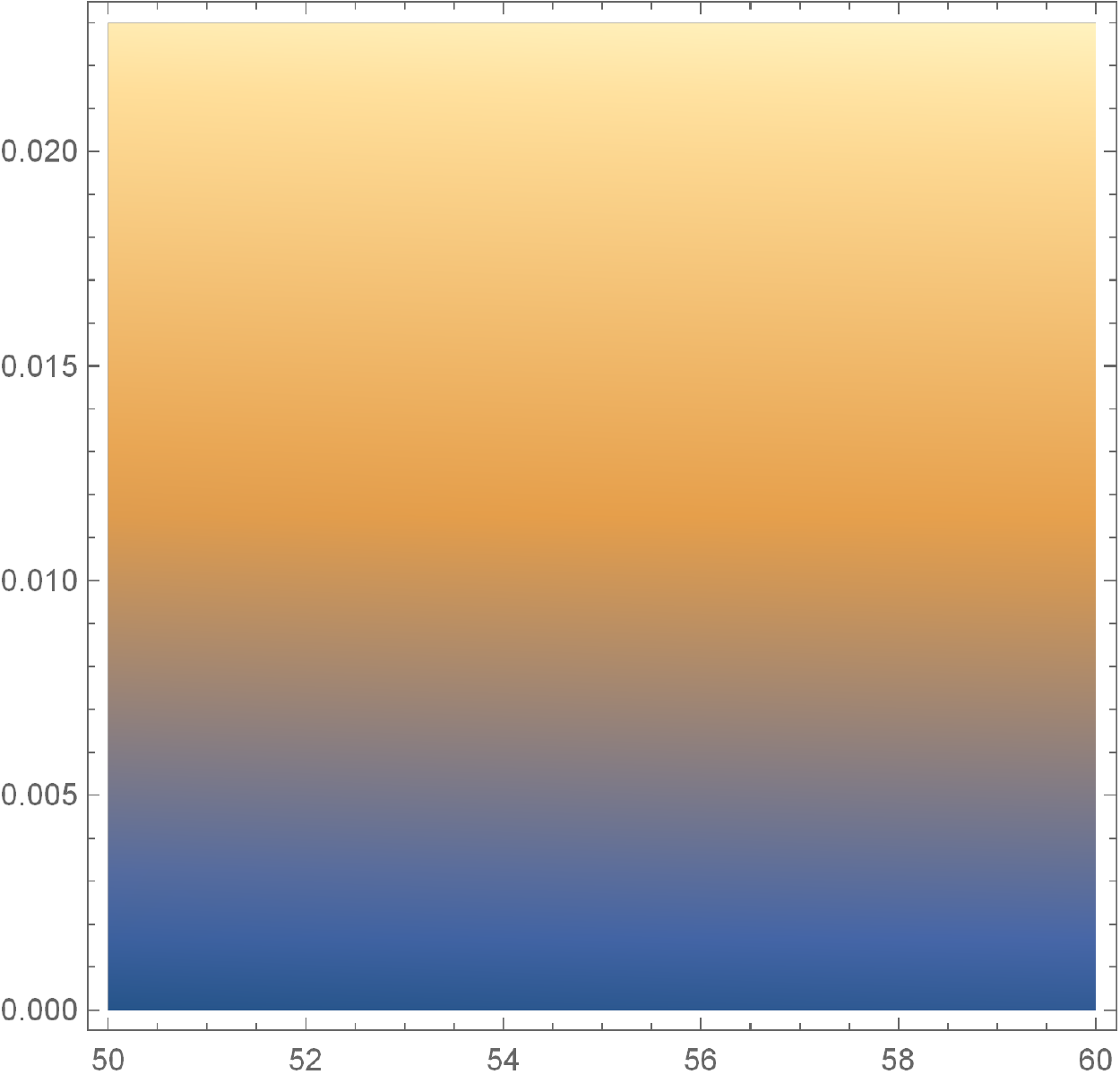}
\includegraphics[height=8.8cm, width=1.05cm ]{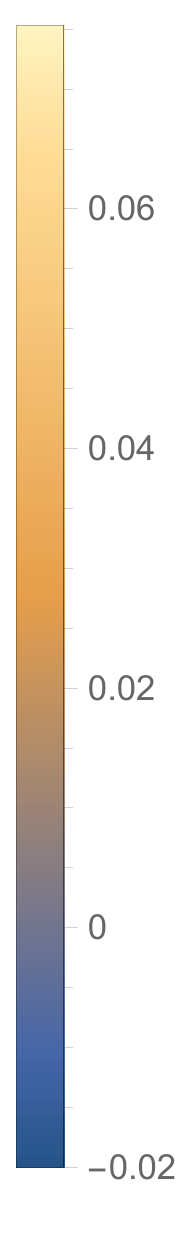}
\vskip -0.4cm
 \caption{{\it Density plot of spectral index (\ref{nchifinal}):} The horizontal line corresponds to e-folding number $N$ whose range is $50\leqslant N\leqslant60$. The vertical line denotes the value of $g_0$ whose range locates from $0.000001$ to $0.023$. The right panel shows that the value of $n_\chi$ matching its corresponding color.}
 \label{spectral index1}
 \end{figure}

Following the standard procedure~\cite{Kofman:1997yn,Kofman:1994rk}, first what we need is the equation of motion (EOM) of curvaton field $\chi$, which can be varied with action (\ref{total action}), one can obtain EOM of $\chi$ field (background field),
\begin{equation}
\ddot{\tilde{\chi}}+\frac{g_0}{M_{P}^{2}}m^{2}\phi^{2}\tilde{\chi}
-\frac{\lambda_{0}\lambda_{1}}{M_{p}}[1-\lambda_{1}\frac{\tilde{\chi}}{M_{P}}])\approx0
,
\label{eom of chi2}
\end{equation}
where changing variable as $\tilde{\chi}=a^{3/2}\chi$ and neglecting the term of $\frac{9}{4}H^{2}\tilde{\chi}+\frac{3}{2}\dot{H}\tilde{\chi}$. In order to solve Eq.~(\ref{eom of chi2}), the solution of background for $\phi$ is mandatory. Its corresponding EOM is derived by,
 \begin{equation}
 \ddot{\phi}+3H\dot{\phi}+\frac{dV(\phi)}{d\phi}=0
 \label{eom of background phi}
 \end{equation}
where $\partial_{\phi}V=\partial_{\phi}(\frac{1}{2}m^{2}\phi^{2}+\frac{1}{2}\frac{g_0}{M_{P}^{2}}\chi^{2}m^{2}\phi^{2})$. Here we define the effective mass of inflaton $m_{eff}^{2}=m^{2}+\frac{g_0}{M_{P}^{2}}\chi^{2}m^{2}$, it leads to $m_{eff}\approx m, \partial_{\phi}V=m_{eff}^{2}\phi$ with $\frac{g_0}{M_{P}^{2}}\chi^{2}\ll1$ illustrating in figure (\ref{spectral index1}). Like deriving Eq. (\ref{eom of chi2}) and neglecting $\frac{9}{4}H^{2}\tilde{\chi}+\frac{3}{2}\dot{H}\tilde{\chi}$. In the later investigation, we will show that this is a rational assumption. Meanwhile, changing variable as $\tilde{\phi}(t)=a^{3/2}\phi(t)$ where $a$ is scale factor and implementing the same trick adopted for deriving Eq.~(\ref{eom of chi2}), Eq.~(\ref{eom of background phi}) finally becomes,
\begin{equation}
\ddot{\tilde{\phi}}+m_{eff}^{2}\tilde{\phi}=0,
\label{eom of background phi1}
\end{equation}
its solution is
\begin{equation}
\phi(t)=a^{-3/2}\cos(m_{eff} t).
\label{solution of phit}
\end{equation}
Using solution (\ref{solution of phit}) to plug into Eq.~(\ref{eom of chi2}) and rearranging this equation with some efforts, this equation becomes,
\begin{equation}
\ddot{\tilde{\chi}}+\frac{g_0m^{2}}{2M_{P}^{2}}\phi_{0}^{2}\tilde{\chi}a^{-3}+\frac{m^{2}g_0}{2M_{P}^{2}}\tilde{\chi}\phi_{0}^{2}a^{-3}\cos(2m_{eff}t)+\frac{\lambda_{0}\lambda_{1}^{2}}{M_{p}^{2}}\tilde{\chi}=\lambda_{0}\lambda_{1}/M_{P},
\label{eom of chi3 with cos}
\end{equation}
Thereafter, one can set $z=m_{eff}t$, Eq.~(\ref{eom of chi3 with cos}) turns into,
\begin{equation}
\tilde{\chi}''+\frac{g_0}{2M_{P}^{2}}\phi_{0}^{2}\tilde{\chi}a^{-3}+\frac{g_0}{2M_{P}^{2}}\phi_{0}^{2}\tilde{\chi}a^{-3}\cos(2z)+\frac{\lambda_{0}\lambda_{1}^{2}}{m_{eff}^{2}M_{p}^{2}}\tilde{\chi}=\lambda_{0}\lambda_{1}/(m_{eff}^{2}M_{P}),
\label{eom of background phi2}
\end{equation}
where we have used the approximation $m_{eff}\approx m$ and $\tilde{\chi}'=\frac{d\tilde{\chi}}{d z}$. Next, we will follow the notation of Ref.~\cite{Shtanov:1994ce}, first the standard form can be written as
\begin{equation}
\tilde{\chi}''+(A_{k}+2q_{k}\cos[2z])\tilde{\chi}=c
\label{eom of background phi3}
\end{equation}
where $c=\lambda_{0}\lambda_{1}/(m_{eff}^{2}M_{P})$, subsequently, one can find the correspondence as follows,
\begin{eqnarray}
A_{k}&=&\frac{g_0\phi_{0}^{2}}{2M_{P}^{2}}a^{-3}+\frac{\lambda_{0}\lambda_{1}^{2}}{m_{eff}^{2}M_{p}^{2}},\\
q_{k}&=&\frac{\phi_{0}^{2}g_0}{2M_{P}^{2}}a^{-3}.
\label{correspondence}
\end{eqnarray}
As mentioned in section \ref{model}, the role of exponential potential $\lambda_0\exp(-\lambda_1\frac{\chi}{M_P})$ is mimicking the evolution of dark energy, thus $\chi$ field will be approached zero and $\lambda_0$ can be determined as the same order with dark energy. Comparing to other terms in Eq.~(\ref{eom of background phi3}), one can approximately set $c\approx 0$. In order to achieve the range of $q$, we analyse the spectral index of curvaton whose detailed calculation is showing in appendix. The formula of spectral index of curvaton is as follows,
\begin{equation}
n_\chi=-2\epsilon_1-\eta_c
\end{equation}
where the definition of $\epsilon_1$ and $\eta_c$ are Eqs.~(\ref{epsilon 1},\ref{curvaton slow roll}) and spectral index is denoted in terms of leading order of slow-roll parameters. Assuming that the slow-roll conditions are applicable for inflaton and curavton. In curvaton scenario, the energy density of inflation is dominant before it decays into curvaton which is also an assumption for curvaton mechanism. Consequently, the value of Hubble parameter is almost determined by inflaton. Afterwards, one adopts the slow-roll approximation of inflation, we can derive the $\epsilon_1=\epsilon_V=\frac{M_P^2}{2}\big(\frac{V'}{V}\big)^2$ where $V'=\frac{\partial_V}{\partial_\phi}$. In light of $N\approx \frac{\phi^2_*}{4M_p^2}$, one easily derives $\epsilon_1=\frac{1}{N}$. As for $\eta=-\frac{2}{3}\frac{V''(\chi)}{H^2}$, we also implement the slow-roll approximation and pay attention $V''(\chi)=\frac{\partial^2 V(\chi)}{\partial \phi^2}$. Subsequently, we remarkably find $\eta_=-4g_0$ which is independent of the model of inflation. Combine these two parts, finally we obtain that
\begin{equation}
n_\chi=-\frac{1}{N}+4g_0,
\label{nchifinal}
\end{equation}
where $N$ is the e-folding number. According to this equation, we can see the varying trends of $n_\chi$ from figure \ref{spectral index1},
 from lasted observational constraint for $n_\chi\approx 0.035$ from \cite{Akrami:2018odb}, the possible value of $g_0\approx 0.013$ being of order $10^{-2}$.

 For spectral index $n_\chi$, its value is almost determined by a large class of inflationary models, in which the e-folding number only depends on kinds of slow roll inflations \cite{Kallosh:2013hoa}. In some sense, $n_\chi$ is independent of inflation. From another perspective, the $g_0$ is practically determined by this constraint. Observing that Eq.~(\ref{correspondence}), $q$ is also determined by $\phi_0$ where it represents the amplitude of inflaton field. At the moment of inflation ends, the value of $\phi_0$ could be of order unity. Then, it will lead to $q$ is of order of unity as well. In light of $q_k\ll1$ and $q_k\geq 1$, equation (\ref{eom of background phi3}) can be treated as narrow resonance and broad resonance, respectively.

\subsection{Narrow resonance}
\label{narrow resonance}

In this case, it corresponds to $q_k\ll1$ which means that $g_0\phi_0^2\ll 2M_P^2$. Ref.~\cite{Shtanov:1994ce} provides a general framework of narrow parametric resonance. The most essential physical quantities are decay rate $\Gamma_\chi$ and number density of resonance $N_{res}$. $\Gamma_\chi$ depicts how much energy has been transferred to curvaton. As for $N_{res}$, it represents the number density of curvaton.

From Ref.~\cite{Shtanov:1994ce}, one can directly find the formulas of $\Gamma_\chi$ and $N_{res}$,
\begin{eqnarray}
N_{res}&=&\sinh^{2}\bigg(\frac{\pi g_{n}^{2}}{2\Gamma_{\chi}\omega_{res}^{3}}\bigg),\\
\Gamma_{\chi}&=&\frac{\pi g_{n}^{2}}{\omega_{res}^{3}}\ln^{-1}\bigg(\frac{32\pi^{2}\rho}{\omega_{res}^{4}}\bigg),
\label{number density}
 \end{eqnarray}
 where $\rho=\frac{1}{2}m^{2}\phi^{2}$ and $n$ denotes the $n$th band of periodic function, here there is only $g_1=g_{1}=\frac{g_0}{2M_{P}^{2}a^{-3}}$ and $\omega_{res}=\frac{1}{2}m_{eff}=\frac{1}{2}\omega$. In order to get efficient production of curvaton, it is requiring that $N_{res}\gg 1$. As a consequence, it is simply concluded that
 $32\pi^{2}\rho\gg\omega_{res}^{4}$ which means that the initial value of $\phi$ is much more lager than its mass (exactly the effective mass) which is consistent with results of chaotic inflation \cite{Linde:1983gd}. Namely, the efficient production requires that inflaton is a large field.

\subsection{Broad resonance}
\label{broad resonance}

 In this case, the theoretical framework \cite{Kofman:1997yn,Kofman:1994rk} provides a general method for studying the broad resonance corresponding to $q_k\geq 1$. Firstly, the standard equation can be written as,
  \begin{equation}
  \ddot{\tilde{\chi}}+\omega^{2}\tilde{\chi}=0,
  \label{eom of chi with broad}
  \end{equation}
where
\begin{equation}
\omega^{2}=+\frac{g_0m^{2}}{2M_{P}^{2}}\phi_{0}^{2}a^{-3}+\frac{m^{2}g_0}{2M_{P}^{2}}\phi_{0}^{2}a^{-3}\cos(2m_{eff}t)+\frac{\lambda_{0}\lambda_{1}^{2}}{M_{p}^{2}}
 ,
 \label{omega in broad}
\end{equation}
and $\dot{\tilde{\chi}}=\frac{d\tilde{\chi}}{dt}$. Although the production of particles are highly non-equilibrium process, one can still use the $WKB$ approximation to analytically solve Eq.~(\ref{omega in broad}), in which if we presume that it is an adiabatic process of a time interval from $t_i$ to $t_{i+1}$ (a very short time interval). Its general solution under WKB approximation,
\begin{equation}
\tilde{\chi}_k^{wkb}\approx\frac{\alpha_{k}}{\sqrt{2\omega}}e^{-i\int^{t}\omega dt}+\frac{\beta_{k}}{\sqrt{2\omega}}e^{i\int^{t}\omega dt}
\label{wkb solution of chi}
\end{equation}
where $\alpha_k$ and $\beta_k$ are constants with adiabatic condition in a short time interval and $\big|\alpha_{k}\big|^{2}-\big|\beta_{k}\big|^{2}=1$. In this short time interval, we can also make an analytical approximation of $\frac{g_0^{2}\phi^{2}}{M_p^2}\approx\frac{m_{eff}^{2}g_0}{2M_{P}^{2}}\phi_{0}^{2}a^{-3}(t_{i+1}-t_{i})^{2}$ and define new variables,
\begin{eqnarray}
\tau^{2}&=&\frac{m_{eff}^{2}g_0}{2M_{P}^{2}}\phi_{0}^{2}a^{-3}(t-t_{j})^{2},\\
\kappa^{2}&=&\frac{\frac{\lambda_{0}\lambda_{1}^{2}}{M_{p}^{2}}}{\frac{m_{eff}^{2}g_0}{2M_{P}^{2}}\phi_{0}^{2}a^{-3}}.
\label{new variables}
\end{eqnarray}
Be armed with these variables, Eq. (\ref{wkb solution of chi}) can be rearranged as
\begin{equation}
\frac{d^{2}\tilde{\chi}_k}{d\tau^{2}}+(\kappa^{2}+\tau^{2})\tilde{\chi_{k}}=0.
\label{wkb solution of chi1}
\end{equation}
There is a Bogoliubov transformation for coefficients $\alpha_k$ and $\beta_k$ of time interval from $t_i$ to $t_{i+1}$.
\begin{equation}
\big(\begin{array}{c}
\alpha_{k}^{i+1}\\
\beta_{k}^{i+1}
\end{array}\big)=\big(\begin{array}{c}
\sqrt{1+e^{-\pi\kappa^{2}}}e^{i\varphi_{k}}\\
-ie^{-\frac{\pi}{2}\kappa^{2}-2i\theta_{k}^{j}}
\end{array}\begin{array}{c}
ie^{-\frac{\pi}{2}\kappa^{2}+2i\theta_{k}^{j}}\\
\sqrt{1+e^{-\pi\kappa^{2}}}e^{-i\varphi_{k}}
\end{array}\big)\big(\begin{array}{c}
\alpha_{k}^{i}\\
\beta_{k}^{i}
\end{array}\big),
\label{bogliubov transformation}
\end{equation}
where $\varphi_{k}=\arg\{\Gamma(\frac{1+i\kappa^{2}}{2})\}+\frac{\kappa^{2}}{2}(1+\ln\frac{2}{\kappa^{2}})$ ($\Gamma$ is special function). Combine with $\big|\alpha_{k}\big|^{2}-\big|\beta_{k}\big|^{2}=1$ and $n_{k}^{i}=\big|\beta_{k}^{i}\big|^{2}$, one can derive this,
\begin{eqnarray}
n_{k}^{i+1}&=&e^{-\pi\kappa^{2}}+(1+2e^{-\pi\kappa^{2}})n_{k}^{i}-2e^{-\frac{\pi}{2}\kappa^{2}}\sqrt{1+e^{-\pi\kappa^{2}}}\sqrt{n_{k}^{i}(1+n_{k}^{i})}\sin\theta_{tot}^{i}\\
\theta_{tot}^{i}&=&2\theta_{k}^{i}-\varphi_{k}+\arg(\alpha_{k}^{i})-\arg(\beta_{k}^{i}).
\label{angular part}
 \end{eqnarray}
 In order to obtain the continuous production, namely the value of $n_{i+1}$ is enhanced comparing to $n_i$. As taking the limit of $i\rightarrow \infty$, we acquire that $n_k\gg1$ satisfied with $\pi\kappa_{n}^{2}\leq1$ which equals to
 \begin{equation}
\frac{\lambda_{0}\lambda_{1}^{2}}{M_{p}^{2}}\leq\frac{\frac{m_{eff}^{2}g_0}{2M_{P}^{2}}\phi_{0}^{2}a^{-3}}{\pi},
 \label{condition}
 \end{equation}
 In light of this observational constraints, the lower bound can be given by (setting $a=1$ and $M_P=1$),
 \begin{equation}
 m_{eff}^2 g_0\phi_0^2\geqslant \frac{2\lambda_{0}\lambda_{1}^{2}}{g_0}.
 \label{upper bound}
 \end{equation}
From this upper bound of $m_{eff}$, one could see that there is a broad range for the validity of $m_{eff}^2\phi_0^2$ since we have not constrained $\lambda_1$ and $\lambda_0$ is of order of cosmological constant, in which it illustrates that there is huge flexibility of our curvaton scenario.

In this section, we analyzed the production of curvaton generated by parametric resonance. During this process, the curvaton field contains two parts: background field and its quantum fluctuation. Since the main contribution for production of curvaton mainly comes via background field, we used the standard procedure to investigate the production of curvaton. From the constraint of coupling $g_0$, the parametric resonance includes the narrow resonance and broad resonance, respectively. In case of narrow resonance, we concluded that the inflaton is a large field comparing to its mass. As for the broad resonance, we have shown that $ m_{eff}^2 g_0\phi_0^2\geqslant \frac{2\lambda_{0}\lambda_{1}^{2}}{g_0}$ which means that there is huge flexibility of our curvaton scenario.

 \section{Power spectrum and Non-Gaussiantiy}
 \label{power spectrum}
 For the production of quantum perturbations, it could be both from the inflaton and curvaton. Following the spirit of traditional curvaton scenario, we assume that the quantum fluctuations from inflation is negligible \cite{Lyth:2001nq}. Therefore, we only concern the dominant contribution of curvature perturbation through curvaton.

Comparing to Ref.~\cite{Jiang:2018uce}, our curvaton corresponds to the entropy field. Once generating curvaton field, it becomes a two field inflationary theory. Therefore, the best way for working is adopting $\delta N$ formalism \cite{Starobinsky:1986fxa,Sasaki:1995aw,Wands:2000dp,Lyth:2004gb,Sasaki:2006kq}.

We work in conformally flat cosmological space-times whose metric is a conformal rescaling of
Minkowski metric,
\begin{equation}
g_{\mu\nu}=a^2(\tau)\eta_{\mu\nu}
\,,\qquad \eta_{\mu\nu}={\rm diag}(-,+,+,+)
\,,
\label{FRLW metric}
\end{equation}
where $\tau$ is conformal time.
In order to obtain the curvaton power spectrum, one ought to solve the operator
equation of motion for the curvaton, which follows from action~(\ref{total action}),
\begin{equation}
\left[\partial_0^2+2\mathscr{H}\partial_0-\nabla^2\right]\hat\chi(x)
 + a^2\hat V_{\,,\chi}(\hat\phi,\hat \chi)=0
\,,
\label{EOM 1}
\end{equation}
where $\hat V_{\,,\chi}\equiv\partial \hat V(\hat\phi,\hat \chi)/\partial \hat\chi$,
 $\mathscr{H}=a^\prime/a$ ($a^\prime=\partial_0 a$) is the conformal Hubble rate,
$\nabla^2\equiv \sum_{i=1}^3\partial_i^2$ and we have neglected the curvaton coupling
to gravitational perturbations, which is for most purposes justified, as the curvaton is
to a good approximation spectator field during inflation.

The field $\hat \chi$ in~(\ref{EOM 1}) satisfies standard canonical quantization relations,
\begin{equation}
[\hat\chi(\tau,\vec{x}),\hat\pi_{\chi}(\tau,\vec{x}')]= i(2\pi)^{3}\delta^{3}(\vec{x}\!-\!\vec{x}')
,\; \;
[\hat\chi(\tau,\vec{x}),\hat\chi(\tau,\vec{x}')]=0
,\; \;
[\hat\pi_{\chi}(\tau,\vec{x}),\hat\pi_{\chi}(\tau,\vec{x}')]=0
\,,
\label{canonical quantisation}
\end{equation}
where $\pi_{\chi}=a^2\chi^\prime$ ($\chi^\prime=\partial_0\chi$)
denotes the curvaton canonical momentum.
Since we are here primarily interested in the curvaton spectrum of free theory,
it suffices to linearise~(\ref{EOM 1}) in small perturbations around the curvaton condensate,
$\langle\hat \chi\rangle \equiv \bar\chi(\eta)$. The procedure of studying
the dynamics of linear curvaton perturbations is standard, details of which can be found
in the Appendix.

By using the $\delta N$ formalism, the power spectrum can be given by \cite{Kawasaki:2011pd,Kobayashi:2012ba}
\begin{equation}
 P_{\zeta *} = \left(\frac{\partial N }{\partial \chi}\frac{H_{*}}{2\pi}\right)^2,
\label{curvature perturbation1}
\end{equation}
where $\partial N / \partial \chi$ is given by
\begin{equation}
  \frac{\partial N }{\partial \chi}
  = \frac13 r_{\rm decay} \frac{1}{1 - X (\chi_{\rm osc})} \left[ \frac{V'(\chi_{\rm osc})}{V (\chi_{\rm osc}) } - \frac{3 X(\chi_{\rm osc})}{\chi_{\rm osc}} \right] \frac{V'(\chi_{\rm osc})}{V'(\chi_\ast)},
\label{dN_dchi}
\end{equation}
with $\chi_{\rm osc} $ and $\chi_*$ being (Einstein frame) field values at the time of the onset of the oscillation and the horizon exit during inflation.
The time of the onset of the curvaton oscillation can be evaluated by
\begin{equation}
\left| \frac{\dot{\chi}}{H \chi} \right| = 1,
\label{onset_osc}
\end{equation}
which can also be written as \cite{Kawasaki:2011pd,Kobayashi:2012ba}
\begin{equation}
H_{\rm osc}^2 = \frac{V'(\chi_{\rm osc})}{c \chi_{\rm osc}}
\label{onset_osc2}
\end{equation}
where $c$ is  given by $ 9/2$ and $5$ when the curvaton begins to oscillate during MD and RD, respectively.
$X(\chi_{\rm osc})$ represents the perturbation generated from the non-uniform onset of the oscillation of the curvaton field, which is written as
\begin{equation}
 X (\chi_{\rm osc})
  = \frac{1}{2 (c - 3)} \left( \frac{\chi_{\rm osc} V'' (\chi_{\rm osc})}{V'(\chi_{\rm osc})} - 1 \right).
\label{X_chi}
\end{equation}
When the potential is quadratic, $X(\chi_{\rm osc})$ vanishes.
$r_{\rm decay}$ roughly corresponds to the fraction of the curvaton energy density to the total one, which is defined as
\begin{equation}
 r_{\rm decay}=\frac{3\bar\rho_{\chi}}{3\bar\rho_{\chi}+4\bar\rho_{\rm rad}}
                =\frac{3\Omega_{\chi}}{3\Omega_{\chi}+4\Omega_{\rm rad}}
\,,\qquad \Omega_{\chi}=\frac{\bar\rho_{\chi}}{\bar\rho_{\chi}+\bar\rho_{\rm rad}}
\,,\quad \Omega_{\rm rad}=\frac{\bar\rho_{\rm rad}}{\bar\rho_{\chi}+\bar\rho_{\rm rad}}.
\label{r: definition}
\end{equation}

Non-linearity parameter $f_{\rm NL}$ is given by \cite{Kawasaki:2011pd,Kobayashi:2012ba}
\begin{align}
f_{\rm NL}
&=  -\frac56 r_{\rm decay} - \frac53 + \frac{5}{2r_{\rm decay}} (1 +   A),
\label{fNL1}
\end{align}
where $A$ is given by
\begin{align}
A &=
 \left[   \frac{V'(\chi_{\rm osc})}{V (\chi_{\rm osc}) } -  \frac{3 X(\chi_{\rm osc})}{\chi_{\rm osc}} \right]^{-1}
 \left[\frac{X'(\chi_{\rm osc})}{1 - X(\chi_{\rm osc}) }
  +  \frac{V^{''}(\chi_{\rm osc})}{V' (\chi_{\rm osc}) } - \left( 1 - X(\chi_{\rm osc}) \right) \frac{V^{''}(\chi_\ast)}{V' (\chi_{\rm osc}) } \right]
 \notag \\
 &
  +\left[  \frac{V'(\chi_{\rm osc})}{V (\chi_{\rm osc}) } -  \frac{3 X(\chi_{\rm osc})}{\chi_{\rm osc}} \right]^{-2}
\left[
 \frac{V^{''}(\chi_{\rm osc})}{V (\chi_{\rm osc}) } - \left( \frac{V'(\chi_{\rm osc})}{V (\chi_{\rm osc}) } \right)^2  -  \frac{3 X'(\chi_{\rm osc})}{\chi_{\rm osc}}
 +  \frac{3 X(\chi_{\rm osc})}{\chi_{\rm osc}^2}  \right] \, .
\label{A}
\end{align}
Here $A$ is characterized by a curvaton with a generic energy potential, in which it experiences a non-uniform onset of its oscillation. Its validity only requires starting a sinusoidal oscillation when satisfying with Eq. (\ref{onset_osc}). Comparing to this method, Ref. \cite{Cai:2010rt}
 introduced a generalized $\delta N$ formulism, the formula of $f_{NL}=\frac{5}{4r}\big(1+\frac{gg"}{g'2}\big)-\frac{5}{3}-\frac{5r}{6}$, where $r=r_{decay}$ and $g\propto \chi$ as the curvaton potential is quadratic at which $g"=0$, meanwhile $g$ depicts the value of curvaton between the Hubble exits and it starts to oscillate. Generically, the information of $g$ corresponds to our $A$.

 In what follows, we will analyse it in our framework showing that the power spectrum $P_\zeta$ and local non-Gaussianity $f_{NL}$ are independent of the potential of inflation. In order to obtain $P_\zeta$ and $f_{NL}$, the key step is to find the relation between $V(\chi_{osc})$ and $V(\chi_*)$, namely we need the relation between $\chi_{osc}$ and $\chi_*$. Through modified Klein-Gordon (KG) equation of $\chi$,
\begin{equation}
\dot{{\chi}}=-\frac{1}{cH}\frac{\partial V(\chi)}{\partial{\chi}},
\label{modified KG eq}
\end{equation}
where $\chi\equiv \chi(t)$ denotes the background field of curvaton. Combine with the potential of curvaton and the definition of $dN=Hdt\rightarrow dt=\frac{dN}{H}$ and integrate both sides of Eq~(\ref{modified KG eq}), then one can obtain,
\begin{eqnarray}
\frac{M_{P}^{2}\log\left(g_0V\chi_{\text{osc}}-\lambda_{0}\lambda_{1}M_{P}+\lambda_{0}\lambda_{1}^{2}\chi_{\text{osc}}\right)}{g_0V+\lambda_{0}\lambda_{1}^{2}}-\frac{M_{P}^{2}\log\left(g_0V\chi_{*}+\lambda_{0}\lambda_{1}^{2}\chi_{*}-\lambda_{0}\lambda_{1}M_{P}\right)}{g_0V+\lambda_{0}\lambda_{1}^{2}}
\nonumber\\=-(\int_{N_{*}}^{N_{end}=0}\frac{dN}{3H^{2}}+\int_{N_{end}}^{N_{osc}}\frac{dN}{cH^{2}}).
\label{relation for chistar and chiosc1}
\end{eqnarray}
So as to achieve the analytic relation between $\chi_{osc}$ and $\chi_*$, in which the contribution of term consisting of $\lambda_0$ can be neglected since its value being of order of dark energy ($\lambda_0\approx 10^{-120}$ as setting $M_P=1$). In light of slow roll conditions $3M_{p}^{2}H^{2}=V$, consequently one derives the relation as follows,
\begin{equation}
\chi_{osc}=\chi_{*}\exp\big[-6g(\frac{1}{c}N_{osc}-\frac{1}{3}N_*)\big],
\label{final relation between chis}
\end{equation}
where $N_{osc}$ and $N_*$ represent the e-folding number during the time of curavton oscillation and the horizon exit, respectively.

Eq.~(\ref{final relation between chis}) tells us that the curvaton approximately decays during inflation and starts to oscillate in RD. Hence, it is a good proxy to the value of curvation during onset of curvaton oscillation, which determines the relation between the $V(\chi_{osc})$ and $V(\chi_*)$. Armed with this relation, we can obtain the statistical properties of curvaton via Eq.~(\ref{curvature perturbation1}) and Eq.~(\ref{fNL1}).

Combine with definition of e-folding number $N=\int dt H$ and slow roll conditions $3M_P^2=V$, in which the potential of inflation is dominant, then one simply derives as $\Delta N=\frac{\phi_*^2}{4M_P^2}-\frac{1}{2}\approx \frac{\phi_*^2}{4M_P^2}$ since the first term is dominant comparing to $\frac{1}{2}$. According to these approximations and $\lambda_0\approx 0$ in Planck units, we remarkably find that
\begin{eqnarray}
P_{\zeta}&=&\frac{H_{*}}{9\pi^{2}}\frac{r^{2}}{\chi_{*}^{2}},\\
f_{NL}&=&-\frac{5r}{6}+\frac{5}{4r}-\frac{5}{3},
\label{power spectrum and fNL}
\end{eqnarray}
which are independent of potential of inflation. In this approximation of $\lambda_0\approx 0$, it corresponds to the case of $A=\frac{1}{2}$ of \cite{Kawasaki:2011pd,Kobayashi:2012ba}, even the relation (\ref{final relation between chis}) between $\chi_{osc}$ and $\chi_*$ is not mandatory. In Ref.~\cite{Cai:2010rt}, $f_{NL}$ will recover Eq. (\ref{power spectrum and fNL}) as curvaton potential is quadratic and there is no nonlinear evolution of curvaton field between the Hubble exits and starts to oscillate, in which this corresponds to curvaton decay at uniform total energy density. Ref. \cite{Cai:2010rt} also studied that the curvaton decay at uniform curavton energy density, in which $f_{NL}=\frac{5}{6}\big(\frac{3(1+w)}{2\tilde{r}}(1+\frac{gg"}{g'^2})+\frac{1-3w}{\tilde{r}}-4\big)$ with $\tilde{r}=\frac{3(1+w)\Omega_\chi}{4+(-1+3w)\Omega_\chi}$ and they discussed that as $w\rightarrow -1$, the amplitude of $f_{NL}$ will be dramatically enhanced in a second inflation. However, this case will not happen in our scenario since the curvaton potential (comparing to exponential part) will be practically disappear after preheating seeing action (\ref{total action}). Meanwhile, the shape of exponential potential is not plateau. Consequently, there is no second inflationary-like process for curvaton. Thus, although we adopt the method of \cite{Cai:2010rt}, it will not influence our main results. In order to elaborate this scenario, we plot $f_{NL}$ and $P_\zeta$.

\begin{figure}[t!]
 \centering
  \includegraphics[width=0.92\textwidth]{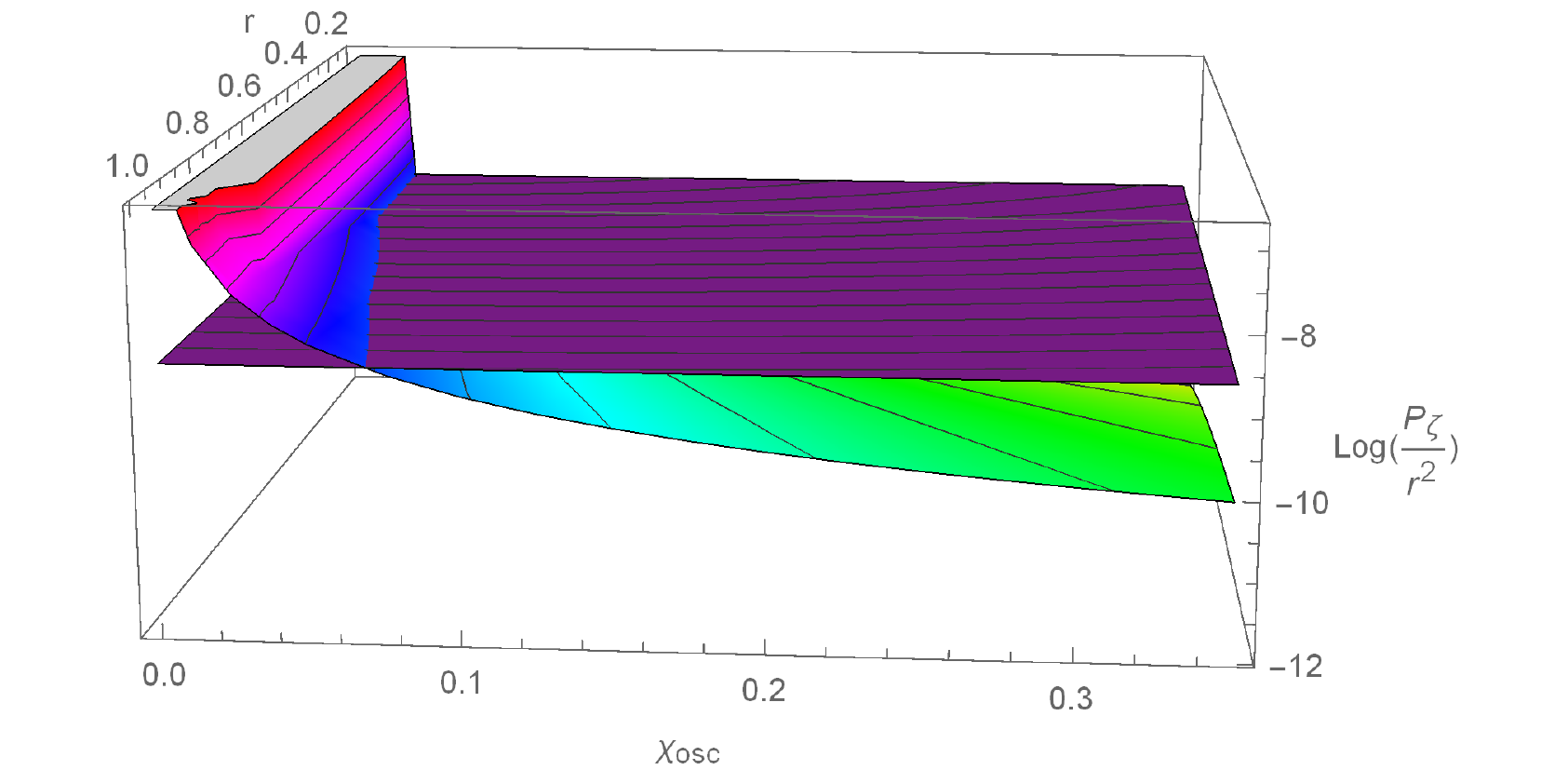}

\vskip -0.4cm
 \caption{{\it Plot of power spectrum:} According to COBE normalization \cite{Akrami:2018odb}, we set $H_I=3.0\times 10^{-5}$ and find $A_s=2.1\times 10^{-9}$ corresponding to the purple plate in this figure, where $r=r_{decay}$.
 \label{figure: power spectrum1}
}
 \end{figure}
\begin{figure}[t!]
 \centering
  \includegraphics[width=0.782\textwidth]{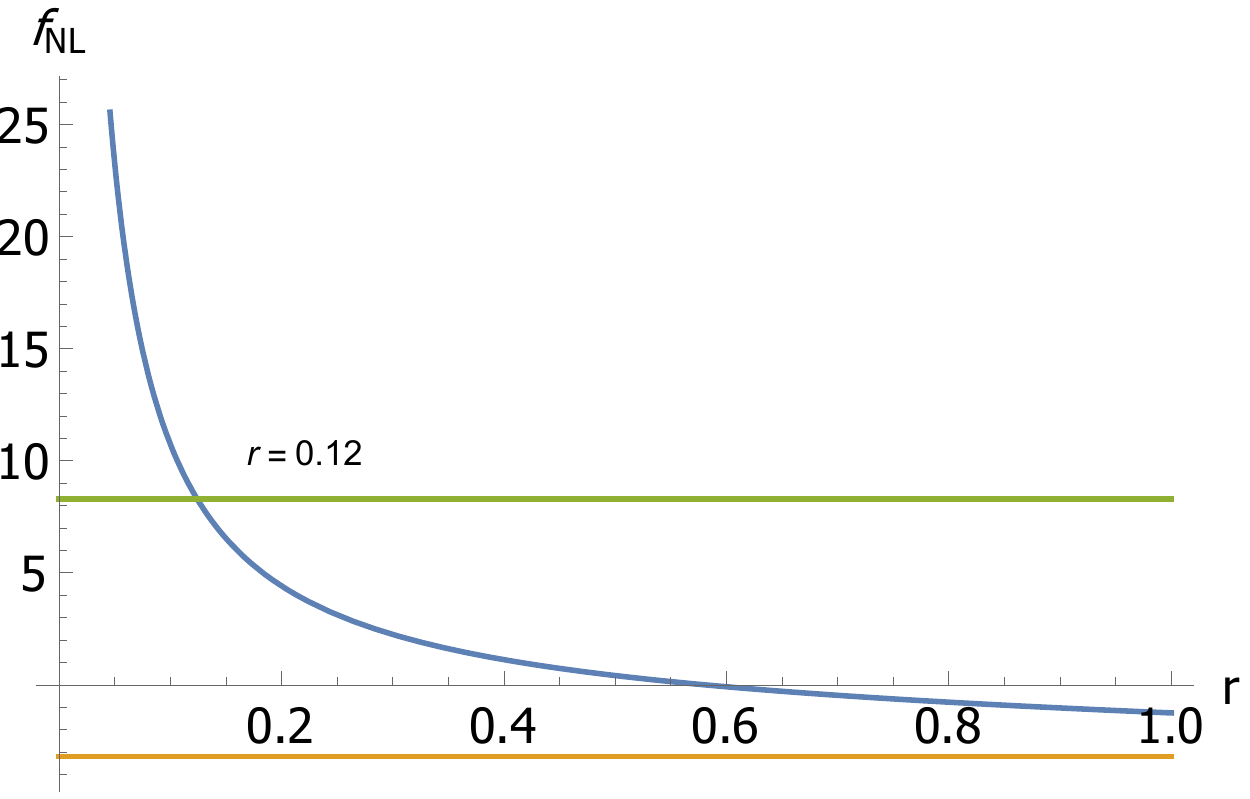}

\vskip -0.4cm
 \caption{{\it Plot of $f_{NL}$:} According to the Planck collaboration constraint on $f_{NL}$ from~\cite{Ade:2015ava}, we show the upper and lower bound of $f_{NL}$, in which we are able to find the lower bound of $r_{decay}\approx 0.12$.
 \label{fnlfinal}
}
 \end{figure}

From figure \ref{figure: power spectrum1}, it tells that the trend of power spectrum with $r_{decay}$ and $\chi_*$, in which it clearly indicates that power spectrum decreases as $\chi_*$ enhances. However, we cannot figure out the range of fraction rate of curvaton $r_{decay}$. In order to compensate this range, the plot of $f_{NL}$ is necessary. From figure \ref{fnlfinal}, it posits the range of $r_{decay}$ of our model.

 Before ending this section, we will give a simple analysis of the decay of curvaton. Explicitly observing that the realization of curvation mechanism happens in the preheating process, meanwhile the main contribution of curvaton's potential is directly coupled to the inflationary potential. Hence for the decay of curvaton, it will be practically  over after preheating (exponential potential is negligible comparing to coupling potential). Moreover, the generation of SM's particles mainly comes via the decay of inflaton since the energy density of inflaton is dominant before preheating. Thus, we only concern that production of particles via the decay of inflaton. When it comes to the precondition of curvaton decay, its effective mass should be larger than the mass of target particle. For instance,
 the creation of Higgs particles, its mass is $125~GeV$ (being of order $10^{-12}$ in Planck units) which means that the effective mass of curvaton should be larger than Higgs mass. However, this process cannot happen since the effective mass for curvaton is $m_\chi=\frac{d^2V(\chi)}{d\chi^2}=$ (being of order of $10^{-120}$) where $V(\chi)=\lambda_0\exp\bigg(-\lambda_1\frac{\chi}{M_P}\bigg)$. The same discussions can be applied for the fermions.
  To the end, the relic of exponential potential will be left for the late Universe.

In this section, we premeditate that power spectrum of and non-local Gaussianity under the framework of $\delta N$ formalism. Taking the appropriate approximation for the calculation, the results imply that it is independent of potential of inflation. From the figure \ref{fnlfinal},~\ref{figure: power spectrum1}, the constraint of $r_{decay}$ can be found, precisely speaking for its lower bound. Once the transferring of energy from inflaton to curvaton occurs, the generation of curvature is a natural process.

\section{Dark energy epoch}
\label{dark energy epoch}

Currently, it is dark energy epoch which responses the accelerated expansion of Universe. However, its origin is still mysterious. During many optional mechanisms, the cosmological constant is the simplest explanation of dark energy  firstly proposed by Einstein \cite{Einstein:1917ce}. It plays a role of dark energy coming from the contribution of James Peebles $\it e.t.c$ \cite{Peebles:2002gy}.

In this scenario, the dark energy will be mimicked by the exponential potential of curvaton in action~(\ref{total action}). From this action, it obviously shows that the effective mass becomes very tiny due the vanishing of inflationary field. Consequently, only the exponential potential of curvaton is dominant, being playing a role of cosmological constant. Current stage is dark epoch dominated by dark energy and the curvaton field is also almost vanishing since it should decay into other particles, especially for the partilces of SM. From lower bound of $\frac{\chi_*}{H_*}=7.0\times 10^ 4$ via \cite{Tenkanen:2019aij} and a good proxy of $H_I=H_*=3.0\times 10^{-5}$ in Planck units, one easily obtains that $\chi_*=2.1$. Then using relation (\ref{final relation between chis}) between $\chi_{osc}$ and $\chi_*$, and relation can approximated to $\chi_{osc}\approx \chi_*\exp(-2g\Delta N)$ where we have used $c\approx 3$ and $\Delta N$ is the variance of e folding number, thus $\Delta N > 60$ and here we set $\Delta N \approx 100$. Meanwhile, the $g_0$ is approximately constraint to $0.01$ from figure \ref{spectral index1}, it yields $\chi_{osc}\approx \chi_*\exp(-2)\approx 0.28$ whose value is located at which the curvaton field starts to oscillate, even it could be smaller due to the decay of curvaton until to being vanishing. From another perspective, there is an extra parameter $\lambda_1$ not determined by observations. Due to the smallness of $\chi_{e}$ denoting the final value after decaying, we have lots of freedom to choose the range of $\lambda_1$ only keeping the $\lambda_1\frac{\chi_e}{M_p}\ll 1$, in which it means that exponential potential is almost to be a constant $\lambda_0$. Afterwards, we retain $\lambda_0\approx 10^{-120}$ in Planck units. This potential naturally plays a role of cosmological constant. Finally, the Universe undergoes the dark epoch.

\section{Conclusion and Outlook}
\label{conclusion}

In this paper, we have constructed a broad classes of curvaton scenarios, in which the effective mass of curvaton is running due to the coupling between curvaton and inflaton. The effective mass of curvaton was proportional to inflationary potential as showing in action~(\ref{total action}). The advantage of this mechanism is as following: $(a)$ The spectral index of curvaton only depends on the e-folding number and coupling coefficient, which means that a large class of inflationary models satisfies with this requirement. $(b)$ Once the curvaton generated by preheating, we calculate the power spectrum and non-linear non-Gaussianity, our calculation shows that these two observables are independent of inflationary potential by neglecting the contribution of exponential potential of curvaton. $(c)$ After the decay of curvaton, the exponential potential of curvaton will be a constant of order of cosmological constant, which may play a role of dark energy. Finally, we construct a large class of curvaton scenarios which is practically independent of model, only requiring that the slow-roll inflationary conditions for inflaton and curvaton. In light of these advantages, we systematically investigate our curvaton mechanism.

Firstly, there was only one field (inflaton) at the very beginning of Universe. Subsequently, energy was transferring from inflaton to curvaton via preheating process. Through section \ref{production of curvaton}
, we have shown that this production contains the background field and quantum fluctuations of curvaton, in which it divided into two cases: narrow resonance and broad resonance. For narrow resonance, using the constraints from the number density of curvaton and decay rate of inflaton (energy transferring from inflaton to curvaton), one naturally concluded that inflationary field was large field (its initial value much larger than its effective mass). As for the broad resonance, the lower bound of $ m_{eff}^2 g_0\phi_0^2\geqslant \frac{2\lambda_{0}\lambda_{1}^{2}}{g_0}$ was given. Once generating the curvaton, we calculated the power spectrum and local Non-Gaussianity. Our results agreed with observational constraints via figure \ref{figure: power spectrum1},\ref{fnlfinal}. Remarkably, we found these results were independent of the inflationary potential albeit effective mass of curvaton was proportional to inflationary potential. This provides us huge freedom to construct the inflationary part.

 Finally, as the decay of curvaton occurred, its field value would be smaller and smaller. From section \ref{dark energy epoch}, we have discussed that the exponential potential would be approaching to be a constant $\lambda_0$ of order of cosmological constant. Therefore, this relic of exponential potential would play a role of cosmological constant from the perspective of phenomenology.

 We will give a outline for the future relevant work. Actually, the curvaton field from preheating process firstly was pioneered investigated under the framework of bounce cosmology \cite{Cai:2011zx}. Afterwards, the same curvaton field can be induced by the preheating process around the nonsingular bounce \cite{Cai:2011ci}. Comparing these works with our curvaton scenario, the difference is for the coupling between the inflaton and curvaton, in which our case is $\chi$ field explicitly couples to the inflationary potential. And our calculation for observable is independent of inflationary potential. Thus, it gives us a hint for realizing curvaton mechanism retaining our virtue under the framework of bounce Universe. However, the nature of inflaton is still mysterious. As Ref. \cite{Traschen:1990sw} mentioned, before a certain moment $t_1$, the expectation value of inflationary field is zero , namely, $\langle\phi^2\rangle=0$. Up to the occurrence of phase transition, $\langle\phi^2\rangle$ becomes non-vanishing, thus the interaction of between $\chi$ and $\phi$ is no longer non-zero, which means that it would generate the mass of curvaton naturally due to the symmetry breaking. Thus, it is automatically consider inflationary field as Higgs field in light of \cite{Enqvist:2012tc,Enqvist:2013gwf}. In our curvaton scenario, we could construct the Higgs field to response the inflation and curvaton is generating the curvature perturbations. Furthermore, the one loop correction can be taken account into considerations under the framework of finite temperature field theory, particularly the effects via temperature to observable. In a near future, one could also use the asymptotic safety to construct the inflationary part \cite{Liu:2018hno}, even we are also interested in study of dark matter constraint via framework of brane world \cite{Li:2017kkj,Li:2018jxy,Xu:2019gzt}.

\section*{Acknowledgements}
LH is grateful for Ai-Chen Li and Hai-qing Zhang of the fruitful discussions and comments for this manuscripts, and thanks for hospitality of Institute of Theoretical Physics in Beijing University of Technology and Beihang University when we begun this project. LH is very precious for his Ph.D supervisor, Prof. Tomislav Prokopec helps the calculation of appendix (forever thanks for his guidance and endless discussions during his whole Phd period). LH is funded by initial started funding of Jishou University. WL is funded by NSFC 1175012.

\section*{Appendix: Linearized curvaton perturbations}
\label{Appendix linearized curvaton perturbations}

In this appendix we recall how to calculate the spectrum of curvaton perturbations during inflation
in the simpleest, tree level (linearized) approximation. On a fixed cosmological gravitational background the curvaton
dynamics is governed by~Eq.~(\ref{EOM 1}),
which is valid provided the curvaton can be regarded as a spectator field,
{\it i.e.} if its energy density is subdominant during inflation.
Assuming this is true and moreover the curvaton perturbations are small, one can linearize~(\ref{EOM 1})
around the background field values, $\bar\chi_E(t)=\langle\hat \chi_E\rangle$, $\bar\phi_E(t)=\langle\hat \phi_E\rangle$ such that,
upon a convenient rescaling and linearization, Eq.~(\ref{EOM 1}) simplifies to,
\begin{equation}
\left(\partial_0^2-\nabla^2+a^2V_E''-\frac{a''}{a}\right)(a\delta \hat\chi_E)=0
\,,
\label{EOM 2}
\end{equation}
where $\delta \hat\chi_E=\hat\chi_E-\langle\hat\chi_E\rangle$,
$V_E'' = \partial^2 V_E(\bar\chi_E,\bar\phi_E)/\partial^2 \bar\chi_E$,
and $a''=d^2a/d\tau^2$.
Since the background~(\ref{FRLW metric}) is invariant under spatial translations, it is natural to assume that the state
respects the same symmetry. In that case one can expand $\delta \hat \chi_E(x)$ in terms of mode functions
$\chi_E(\tau,k)$ and $\chi^*_E(\tau,k)$ as follows,
 \begin{equation}
\delta \hat{\chi}_E(\tau,\vec x)=\int\frac{d^{3}k}{(2\pi)^{3}}e^{\imath \vec{k}\cdot \vec{x}}
    \left[\chi_E(\tau,k)\hat{a}(\vec{k}\,)+\chi_E^*(\tau,k)\hat{a}^\dag(-\vec{k}\,)\right]
\,,
\label{Fourier modes for curvaton}
\end{equation}
where $k=\|\vec{k}\|$, $\hat{a}(\vec{k}\,)$ is the particle annihilation operator
that annihilates the vacuum $|\Omega\rangle$,
$\hat{a}(\vec{k})|\Omega\rangle=0$, and $\hat{a}^+(\vec{k}\,)$ is the particle creation operator
that creates one quantum of momentum $\vec k$. These operators obey,
\begin{equation}
 [\hat a(\vec{k}\,),\hat a^+(\vec{k}'\,)]= (2\pi)^{3}\delta^{3}(\vec{k}\!-\!\vec{k}'\,)
\,,\; \;
[\hat a(\tau,\vec{k}\,),\hat a(\tau,\vec{k}'\,)]=0
\,,\; \;
[\hat a^+(\tau,\vec{k}\,),\hat a^+(\tau,\vec{k}'\,)]=0
\,.
\label{canonical quantisation: a and a dagger}
\end{equation}
From~(\ref{EOM 2}) one can see that the mode function $\chi_E(\tau,k)$ satisfies
the following differential equation,
 \begin{equation}
\left(\frac{d^2}{d\tau^2}+k^2 + a^2V_E''-\frac{a''}{a} \right)[a\chi_E(\tau,k)] = 0
\,.
\label{EOM 3}
\end{equation}
The last term in~(\ref{EOM 3}) can be written as,
\begin{equation}
\frac{a''}{a} = {\cal H}_E^2\left(1+\frac{{\cal H}_E'}{{\cal H}_E^2}\right)
            = {\cal H}_E^2\left(2-\epsilon_1\right)
\label{a''/a}
\end{equation}
\begin{equation}
 \epsilon_1= -\frac{\dot H_E}{H_E^2} = 1-\frac{{\cal H}_E'}{{\cal H}_E^2}
\label{epsilon 1}
\end{equation}
is the principal slow roll parameter. The conformal Hubble parameter ${\cal H}_E$ can be expressed
in terms of conformal time and a power series in slow roll parameters as follows,
\begin{equation}
{\cal H}_E =-\frac{1}{\tau}\left[1+\epsilon_1+\epsilon_1(\epsilon_1+\epsilon_2)+{\cal O}(\epsilon_i^3)\right]
\,.
\label{conformal Hubble}
\end{equation}
Taking account of these relations and calculating to the second order in slow roll parameters,
Eq.~(\ref{EOM 3}) becomes
 \begin{equation}
\left(\frac{d^2}{d\tau^2}+k^2 -\frac{1}{\tau^2}\left[2\!+\!3\epsilon_1\!+\!4\epsilon_1(\epsilon_1\!+\!\epsilon_2)
 +\frac32 (1\!+\!2\epsilon_1\!
     )\eta_c+{\cal O}(\epsilon_i^3,\eta_c\epsilon_j^2)\right]\right)[a\chi_E(\tau,k)] = 0
\,,
\label{EOM 3b}
\end{equation}
where we have introduced the {\it principal curvaton slow roll parameter} $\eta_c$ and
the second slow roll parameter $\epsilon_2$ as,
\begin{equation}
\eta_c = -\frac23\frac{V_E''}{H_E^2}
\,,\qquad \epsilon_2=\frac{\dot\epsilon_1}{\epsilon_1H}
\,.
\label{curvaton slow roll}
\end{equation}
Assuming the term in~(\ref{EOM 3b}) that multiplies $1/\tau^2$ varies adiabatically in time,
Eq.~(\ref{EOM 3b}) can be solved  in terms of Hankel functions. The fundamental solutions are given by,
\begin{equation}
\psi(\tau,k)=\frac{1}{a}\sqrt{\frac{-\pi\tau}{4}}H_\nu^{(1)}(-k\tau),
\qquad
\psi^*(\tau,k)=\frac{1}{a}\sqrt{\frac{-\pi\tau}{4}}H_\nu^{(2)}(-k\tau)
\,,
\label{Hankel function}
\end{equation}
their Wronskian normalization is,
\begin{equation}
W[\psi(\tau,k),\psi^*(\tau,k)]=\frac{\imath}{a^2}
\,.
\label{Wronskian}
\end{equation}
and the index reads,
\begin{equation}
\nu^2=\frac94+3\epsilon_1+\frac32\eta_c+\epsilon_1(4\epsilon_1\!+\!4\epsilon_2+3\eta_c)
 \;\Longrightarrow\; \nu\simeq \frac32 +\epsilon_1+\frac{1}{2}\eta_c
          +\frac13\epsilon_1\left(3\epsilon_1+4\epsilon_2+3\eta_c\right) +{\cal O}(\epsilon_i^3,\eta_c\epsilon_i^2)
\,.
\label{nu2 and nu}
\end{equation}
The general mode consistent with spatial homogeneity and isotropy is then,
\begin{equation}
 \chi_E(\tau,k)= \alpha(k) \psi(\tau,k)+\beta(k)\psi^*(\tau,k)
\,,\qquad  |\alpha(k)|^2-|\beta(k)|^2=1
\,.
\label{general mode function}
\end{equation}
A standard Bunch-Davies choice of the vacuum amounts to $\alpha(k)=1$ and $\beta(k)=0$,
which is what we assume throughout this work.

 The corresponding power spectrum and  spectral index are defined by,
 \begin{equation}
P_{\chi_E}(\tau,k)=\frac{k^3}{2\pi^2}|\chi_E|^2=P_{\chi_E*}\left(\frac{k}{k_*}\right)^{n_{\chi_E}}
\,.
\label{power spectrum and index}
\end{equation}
By making use of~(\ref{general mode function}) and~(\ref{Hankel function})
in~(\ref{power spectrum and index}) one obtains,
 \begin{equation}
P_{\chi_E}(k,\tau)=\frac{1}{a^2}\frac{k^3|\tau|}{8\pi}|H_\nu^{(1)}(-k\tau)|^2
\,.
\label{power spectrum: curvaton}
\end{equation}
We are particularly interested in super-Hubble scales, where $|k\tau|\ll 1$, and
the Hankel functions of the first kind,
 \begin{equation}
H_\nu^{(1)}(-k\tau)=\frac{1}{\sin(\pi\nu)}\left[e^{i\pi\nu}J_\nu(-k\tau)-\imath J_{-\nu}(-k\tau)\right]
~~~~(|\arg[-k\tau]| < \pi)
\,
\label{Hankel function 1}
\end{equation}
can be expanded as,
\begin{equation}
H_\nu^{(1)}(-k\tau)=\frac{1}{\pi}\left[-e^{i\pi\nu}\Gamma(-\nu)\left(\frac{-k\tau}{2}\right)^\nu-\imath \Gamma(\nu)\left(\frac{-k\tau}{2}\right)^{-\nu}\right]
+{\cal O}\left(|k\tau|^{\nu+2},|k\tau|^{-\nu+2}\right)
\,.
\label{Hankel function expansion}
\end{equation}
Since $\nu > 0$, the second term of Eq.~(\ref{Hankel function expansion}) dominates
and we arrive at the curvaton power spectrum on super-Hubble scales,
  \begin{equation}
P_{\chi_E}(\tau,k)=\frac{H_E^2\Gamma^2(\nu)}{\pi^3[1+\epsilon_1+\epsilon_1(\epsilon_1+\epsilon_2)]^2}
            \left[\frac{k[1+\epsilon_1+\epsilon_1(\epsilon_1+\epsilon_2)]}{2H_Ea}\right]^{n_\chi}
\,,
\label{final power spectrum}
\end{equation}
with $\nu$ given in~(\ref{nu2 and nu}),
\begin{equation}
n_\chi = 3-2\nu = -2\epsilon_1-\eta_c
          -\frac23\epsilon_1\left(3\epsilon_1+4\epsilon_2+3\eta_c\right) +{\cal O}(\epsilon_i^3,\eta_c\epsilon_i^2)
\label{spectral index}
\end{equation}
 where we have used,
$-k\tau\approx k[1+\epsilon_1+\epsilon_1(\epsilon_1+\epsilon_2)]/(H_Ea)$, see~(\ref{conformal Hubble}).
From~(\ref{final power spectrum}) one can easily read off the spectrum amplitude $P_{\chi_E*}$.
To linear order in the slow roll parameters it reads,
 \begin{equation}
P_{\chi_E* }(t,k_*)=\frac{H_E^2}{4\pi^2}\Big[1\!-\!2\epsilon_1\!+\!\frac23(2\epsilon_1\!+\!\eta_c)\psi(3/2)\Big]
\exp\Big[\!-\!n_\chi\Big(N+\ln\frac{2H_E(1\!-\!\epsilon_1)}{k_*}\Big)\Big]
\,,
\label{power spectrum and index:2}
\end{equation}
where $\psi(3/2)=2-\gamma_E-2\ln(2)\approx 0.0367$ is the di-gamma function of $3/2$
and $H^2\simeq H_0^2e^{-2\epsilon_1N}$.
We see that the amplitude $P_{\chi_E* }$ depends weekly on time. For example, for a red-tilted spectrum
for which $n_\chi<0$,
$P_{\chi_E* }$ grows exponentially with the number of e-foldings $N=\ln(a)$, {\it e.g.}
for $n_\chi\simeq -0.04$ and $\epsilon_1=0.01$, $P_{\chi_E* }$ grows about $2\%$ per e-folding.

\bibliography{mybibfile}

\end{document}